\documentstyle[prl,aps]{revtex}        
\begin{document}
\draft
\twocolumn
\title{Explanation of
the violation of Bell's inequality by maintaining Einstein separability} 
\author{Gyula Bene}
\address{Institute for Solid State Physics, E\"otv\"os University,
M\'uzeum krt. 6-8, H-1088 Budapest, Hungary}
\date{\today}
\maketitle

\begin{abstract}
A new theory is proposed that offers a 
consistent conceptual basis for nonrelativistic
quantum  mechanics. The violation of Bell's inequality is explained
 by maintaining realism, inductive inference
and Einstein separability.  
\end{abstract}
\pacs{03.65.Bz}
\narrowtext

Despite the successes of quantum mechanics its basic concepts,
especially the measurement and the collapse of the wave
function \cite{von Neumann}  has remained
unclear and controversial up to these days. From the theoretical 
point of view, the most serious
problem  is probably
the violation of Bell's inequality\cite{hid}, as it is believed
to be the proof that 
Einstein separability is violated in Nature. Einstein separability 
is an obvious
physical requirement stating that separated systems (i.e., which are prevented
from any interaction with each other) cannot
influence each other.

As a brief reminder, let us think of two spin-half particles 
in an entangled state, 
which is due to some previous interaction between them.
Suppose that the particles are already separated so much that
they can no longer interact with each other. 
Imagine that we perform spin measurements in different directions
 on each
of the two separated particles. 
It is natural to assume that any
correlation between the results of the measurements performed
on the different particles can come only from the previous
interaction which created the entangled state. One may also assume that
there are some
stable properties attached to each system, so that these
properties 'store' the correlation after the systems have become
separated. Using these assumptions one may derive that the
 correlations cannot be
arbitrary but must satisfy a certain inequality. This is Bell's
inequality. The correlations may be calculated quantum
mechanically, and the quantum prediction {\em does not} always 
satisfy Bell's inequality. Correlations are measurable
quantities, and experiments\cite{exp} have proved the correctness of 
the  quantum prediction and thus the violation of Bell's
inequality.

Most people seem to believe that the above
result implies that separated systems can influence each
other. This belief is based on the careful analysis of the
above sketched derivation of Bell's inequality. It turns out that
this derivation
is completely independent of quantum mechanics, and it is
based on a few very fundamental assumptions\cite{Esp}: realism,
inductive inference and Einstein separability. Realism and
inductive inference are so important in physics that we certainly 
do not want 
to give up them. The usual conclusion is that
Einstein separability is violated. 

Nevertheless, we maintain that such a conclusion is 
physically unacceptable. The principle of Einstein
separability has served us well in every branch of physics,
even in quantum physics, including the most sophisticated
quantum field theories. The only way out can be if there is some further,
independent and hidden assumption, which seems to us 
obvious, but which is not valid in quantum mechanics.

In the present letter it will be shown that it is indeed the
case.  One may reinterpret the meaning and the interrelations of the
quantum states such a manner that 
the violation of Bell's inequality gains a natural
explanation without giving up realism, inductive inference or
Einstein separability. The hidden, not allowed assumption mentioned above
is connected to the fact that in the new theory 
it may happen that different states, although individually exist,
cannot be compared.  
In case of the violation of Bell's inequality it turns out that
the states of the measuring 
devices and those which 'store' the correlations are not
comparable as any attempt for a comparison changes the correlations themselves.
Therefore, the usual picture about stable properties 
which 'store' the correlations and are comparable in principle 
at any time with anything
does not apply,
although the correlations may be attributed exclusively
to the 'common past' (previous interaction) of the particles.

To begin with, let us consider a simple example, namely, an idealized measurement
of an $\hat S_z$ spin component of some spin-$\frac{1}{2}$ 
particle. Be the particle $P$ initially in the state
$
\alpha |\uparrow>+\beta|\downarrow>$,
where $|\alpha|^2+|\beta|^2=1$ and the states $|\uparrow>$ and $|\downarrow>$ are the
eigenstates of $\hat S_z$ corresponding to the eigenvalues
$\frac{\hbar}{2}$ and $-\frac{\hbar}{2}$, respectively.
The dynamics of the
measurement is given by the relations
$|\uparrow>|m_0>\; \rightarrow \; |\uparrow>|m_{\uparrow}>$ and  
$|\downarrow>|m_0>\; \rightarrow \; |\downarrow>
|m_{\downarrow}>$, where $|m_0>$, $|m_{\uparrow}>$ and $|m_{\downarrow}>$ stand for states 
of the measuring device $M$. 
The linearity of the Schr\"odinger equation
implies that the measurement process can be 
written as
\begin{eqnarray}
(\alpha |\uparrow>+\beta |\downarrow>)|m_0> \mbox{\hspace{3cm}}\nonumber\\
\rightarrow \; |\Psi>=\alpha |\uparrow>|m_{\uparrow}>
+\beta |\downarrow>|m_{\downarrow}>\quad.
 \label{u2}
\end{eqnarray}
Let us consider now the state of the measuring device $M$ after the measurement. As the
combined system $P+M$ is in an entangled state, the 
measuring device has no own wave function and may be described
by the {\em reduced density matrix}\cite{Landau} 
\begin{eqnarray}
\hat \rho_M=Tr_P\left(|\Psi><\Psi|\right)\mbox{\hspace{3cm}}\nonumber\\
=|m_{\uparrow}>|\alpha|^2<m_{\uparrow}|
+ |m_{\downarrow}>|\beta|^2<m_{\downarrow}|\quad,\label{u3}
\end{eqnarray}
where $Tr_P$ stands for the trace operation in the Hilbert
space of the particle $P$. Nevertheless, if we look at the measuring device,
we certainly see that either $\frac{\hbar}{2}$ or $-\frac{\hbar}{2}$ spin component
has been measured, that correspond to the states $|m_{\uparrow}> $ and $|m_{\downarrow}>$,
respectively. These are obviously not the same as the state (\ref{u3}).
 Why do we get different states?
According to orthodox quantum mechanics,
one may argue as follows. The reduced density matrix $\hat \rho_M$ 
has been calculated from the state $|\Psi>$ (cf. Eq.(\ref{u2}))
of the whole system $P+M$. A state is a result of a measurement
(the preparation), so we may describe $M$ by $\hat \rho_M$ if we have
gained our information about $M$ from a measurement done on $P+M$. 
On the other hand, looking at the measuring device directly
is equivalent with a measurement done directly on $M$.
In this case $M$ is described by either $|m_{\uparrow}> $ or $|m_{\downarrow}>$.
We may conclude that performing  measurements on 
different systems (each containing the system we want to decribe)
gives rise to different descriptions 
(in terms of different states).
Let us call the system which has been measured (it is $P+M$
in the first case and $M$ in the second case) the {\em quantum
reference system}. Using this terminology, we may tell that
we make a measurement on the quantum reference system $R$, thus we prepare
its state $|\psi_R>$ and using this information we calculate
the state $\hat \rho_S(R)=Tr_{R\setminus S} |\psi_R><\psi_R|$ 
of a subsystem $S$. We shall call
$\hat \rho_S(R)$  the state of $S$ with respect to $R$. 
Obviously $\hat \rho_R(R)=|\psi_R><\psi_R|$, thus $|\psi_R>$
may be identified with the state of the system $R$ 
with respect to itself. For brevity we shall call this the internal state of $R$.

Let us emphasize that up to now, despite of the new terminology,
there is nothing new in the discussion. We have merely 
considered some rather elementary consequences of basic quantum mechanics. 

Let us return now to the question why the state of
the system $S$ (i.e., $\hat \rho_S(R)$) depends
 on the choice of the quantum reference
system $R$. In the spirit of the Copenhagen interpretation
one would answer that in quantum mechanics measurements
unavoidably disturb the systems, therefore, if we perform
measurements on  different surroundings $R$, this disturbance is
different, and this is reflected in the $R$-dependence of $\hat \rho_S(R)$.
Nevertheless, this argument is not compelling. 
At this decisive point we leave the traditional framework of quantum mechanics
and 
assume that the states of the systems have already existed before the
measurements, and that there may exist measurements which do not change
these states.  Then the $R$-dependence
of $\hat \rho_S(R)$ becomes an inherent property of quantum mechanics.

The meaning of the quantum reference systems is now analogous
to the classical coordinate systems. Choosing a
classical coordinate system means that we imagine
what we would experience if we were there. Similarly,
choosing a quantum reference system $R$ means that we
imagine what we would experience if we did a measurement on
$R$ that does not disturbe $\hat \rho_R(R)=|\psi_R><\psi_R|$. 
In order to see that such a measurement exists, consider an
operator $\hat A$ (which acts on the Hilbert space of $R$) whose
eigenstates include $|\psi_R>$. The measurement of $\hat A$
will not disturbe $|\psi_R>$. Let us emphasize that the
possibility of nondisturbing measurements is an expression
of realism: the state $\hat \rho_R(R)$ exists independently
whether we measure it or not.

As the dependence of $\hat \rho_S(R)$ on $R$
is a fundamental property now, one has to specify 
the relation of the different states in  terms of suitable
postulates:

{\bf Postulate A.\em  If the reference system $R=I$ is an isolated one
\footnote{An isolated system is 
such a system that has not been interacting 
with the outside world. A closed system
is such a system that is not interacting with any other
system at the given instant of time 
(but might have interacted in the past).}
then the state $\hat \rho_S(I)$ commutes with the
internal state $\hat \rho_S(S)$.}

This means that the internal state of $S$ coincides with
one of the eigenstates of $\hat \rho_S(I)$. Therefore, we shall call 
the eigenstates of
$\hat \rho_S(I)$ the possible internal states of $S$
provided that the reference system $I$ is an
 isolated one.

{\bf Postulate B.\em  The result of a measurement is contained 
unambigously in the internal state of the measuring device.}

{\bf Postulate C. \em If there are $n$ ($n=1,\;2,\;3,\;...$) disjointed physical systems, 
denoted by
\hfill\break
$S_1, S_2, ... S_n$, all contained in the isolated reference 
system $I$ and 
having the 
possible internal states
$|\phi_{S_1,j}>,...,|\phi_{S_n,j}>$, respectively, 
then the joint
probability that $|\phi_{S_i,j_i}>$ 
coincides with the internal state of $S_i$ ($i=1,..n$)
is given by
\begin{eqnarray}
P(S_1,j_1,...,S_n,j_n)\mbox{\hspace{3.5cm}}\nonumber\\
=Tr_{S_1+...+S_n} [\hat \pi_{S_1,j_1} 
...\hat \pi_{S_n,j_n}\hat \rho_{S_1+...+S_n}(I)],\label{u5}
\end{eqnarray}
where $\hat \pi_{S_i,j_i}=|\phi_{S_i,j_i}><\phi_{S_i,j_i}|$.}

Furthermore, the time dependent Schr\"odinger equation remains valid for closed systems.
Note that in the present theory there is no collapse or reduction of the wave function,
and every conclusion should be drawn by using the above rules.

Let us consider now a two-particle system $P_1+P_2$ consisting of
the separated spin-half particles $P_1$ and $P_2$. The initial
internal state of the two-particle system be
\begin{eqnarray}
\sum_j c_j
|\phi_{P_1,j}>|\phi_{P_2,j}>\label{u14}
\end{eqnarray}
where $c_1=a$, $c_2=-b$ (certainly $|a|^2+|b|^2=1$), $
|\phi_{P_1,1}>=|1,\uparrow>$, $
|\phi_{P_1,2}>=|1,\downarrow>$, $
|\phi_{P_2,1}>=|2,\downarrow>$, $
|\phi_{P_2,2}>=|2,\uparrow>$.
When the two-particle system is in the state (\ref{u14}),
there are strong correlations
between the states $\hat \rho_{P_1}(P_1)$\hfill\break 
$=|\psi_{P_1}><\psi_{P_1}|$ 
and $\hat \rho_{P_2}(P_2)=|\psi_{P_2}><\psi_{P_2}|$. 
Provided that the system $P_1+P_2$
is initially isolated, applying {\bf Postulate C} we obtain that the
probability that $|\psi_{P_1}>=|\phi_{P_1,j}>$ and 
$|\psi_{P_2}>=|\phi_{P_2,k}>$ is
$
P(P_1,j,P_2,k)=|c_j|^2\delta_{j,k}
$.

Let us consider now a typical experimental situation, 
when measurements on both
particles are performed. Before the measurements the internal
state of the isolated system $P_1+M_1+P_2+M_2$ ($P_1,P_2$
standing for the particles and $M_1,M_2$ for the measuring
devices, respectively) is given by
$
\left(\sum_j c_j
|\phi_{P_1,j}>|\phi_{P_2,j}>\right)|m^{(1)}_0> |m^{(2)}_0>$,
while it is
\begin{eqnarray}
\sum_j c_j
\hat U_t(P_1+M_1)\left(|\phi_{P_1,j}>|m^{(1)}_0>\right)\nonumber\\
\times\hat U_t(P_2+M_2)\left(|\phi_{P_2,j}>|m^{(2)}_0>\right) \quad,\label{u15}
\end{eqnarray}
with a time $t$ later, i.e. during and after the measurements. Here 
$\hat U_t(P_i+M_i)$ ($i=1,2$) stands for the unitary time evolution operator
of the closed system $P_i+M_i$.

Eq.(\ref{u15}) implies (according to {\bf Postulate A}) that the internal states of 
the closed systems $P_1+M_1$ and $P_2+M_2$
evolve unitarily and do not influence each
other. This time evolution can be given explicitly through the 
relations
\begin{eqnarray}
|\xi(P_i,j)>|m^{(i)}_0>\;\rightarrow \;|\xi(P_i,j)>|m^{(i)}_j>\quad,
\label{u16}
\end{eqnarray}
where $i,j=1,2$ and $|\xi(P_i,j)>$ is the $j$-th eigenstate of the 
spin measured 
on the $i$-th particle along an axis $z^{(i)}$ which closes an angle
$\vartheta_i$ with the original $z$ direction. The time evolution of 
the internal state of the closed systems $P_i+M_i$ is given explicitly by
$
|\psi_{P_i}>|m_0^{(i)}>\;\rightarrow \;
\sum_j <\xi(P_i,j)|\psi_{P_i}>|\xi(P_i,j)>|m_j^{(i)}>
$.
As we see, the $i$-th measurement process 
is completely determined by the initial internal state of the
particle $P_i$. Therefore, any correlation between
the measurements may only stem from the initial correlation
of the internal states of the particles.

The calculation of the state $\hat \rho_{M_1}(M_1)$ (which corresponds to the
measured value, cf. {\bf Postulate B}) needs the state of the whole isolated system
$P_1+P_2+M_1+M_2$.
Using Eq.(\ref{u16}) the final state (\ref{u15}) may be written as
$
\sum_{j,k}\left(
\sum_l c_l<\xi(P_1,j)|\phi_{P_1,l}><\xi(P_2,k)|\phi_{P_2,l}>
\right)
|m^{(1)}_j>|m^{(2)}_k>|\xi(P_1,j)>|\xi(P_2,k)>$.
Direct calculation shows that
$
\hat \rho_{M_1}(P_1+P_2+M_1+M_2)=\sum_j\left(\sum_l
|c_l|^2 |<\xi(P_1,j)|\phi_{P_1,l}>|^2\right)|m^{(1)}_j><m^{(1)}_j|
$.
Note that it is independent of the second measurement.
According to {\bf Postulate A}  $|\psi_{M_1}>$ is one of the
$|m^{(1)}_j>$-s. 
The probability to observe the $j$-th result (up or
down spin in a chosen direction) is
$
P(M_1,j)=\sum_l |c_l|^2 |<\xi(P_1,j)|\phi_{P_1,l}>|^2$.
This may be interpreted in conventional terms: $|c_l|^2$ is
the probability that $|\psi_{P_1}>=|\phi_{P_1,l}>$,
and $|<\xi(P_1,j)|\phi_{P_1,l}>|^2$ is the conditional
probability that one gets the $j$-th result if $|\psi_{P_1}>=|\phi_{P_1,l}>$.
Thus we see that the initial internal state of $P_1$ determines
the outcome of the first measurement in the usual 
probabilistic sense.  
But doesn't it mean that the internal states of $P_1$ and $P_2$
play the role of local hidden variables? No, because hidden variables
are thought to be comparable with the results of the measurements
so that their joint probability may be defined, while in our theory
there is no way to define the joint probability
$P(P_1,l_1,P_2,l_2,(0);M_1,j,M_2,k,(t))$, i.e., the probability that initially
$|\psi_{P_1}>=|\phi_{P_1,l_1}>$ and $|\psi_{P_2}>=|\phi_{P_2,l_2}>$
{\em and} finally $|\psi_{M_1}>=|m^{(1)}_j>$ 
and $|\psi_{M_2}>=|m^{(2)}_k>$. Intuitively we would write
\begin{eqnarray}
P(P_1,l_1,P_2,l_2,(0);M_1,j,M_2,k,(t))\mbox{\hspace{1cm}}\nonumber\\
=|c_{l_1}|^2\delta_{l_1,l_2}|<\xi(P_1,j)|\phi_{P_1,l_1}>|^2\mbox{\hspace{1cm}}\label{u18}\\
\times|<\xi(P_2,k)|\phi_{P_2,l_2}>|^2,\mbox{\hspace{2.3cm}}\nonumber
\end{eqnarray}
as $|c_{l_1}|^2\delta_{l_1,l_2}$ is
the joint probability that $|\psi_{P_1}>=|\phi_{P_1,l}>$ 
and $|\psi_{P_2}>=|\phi_{P_2,l}>$, and $|<\xi(P_i,j)|\phi_{P_i,l_i}>|^2$ is the conditional
probability that one gets the $j$-th result in the $i$-th
measurement if initially $|\psi_{P_i}>=|\phi_{P_i,l_i}>$ ($i=1,2$).
Certainly the existence of such a joint probability would immediately imply the
validity of Bell's inequality, thus it is absolutely important 
to understand why this probability does not exist.

Let us mention, first of all, that using  {\bf  Postulate C} for $n=2$, we may calculate the correlation
between the measurements, i.e., the joint probability 
that $|\psi_{M_1}>=|m^{(1)}_j>$ {\em and} $|\psi_{M_2}>=|m^{(2)}_k>$.
We obtain
\begin{eqnarray}
P(M_1,j,M_2,k)\mbox{\hspace{5cm}}\nonumber\\
=\left|\sum_l c_l<\xi(P_1,j)|\phi_{P_1,l}><\xi(P_2,k)|\phi_{P_2,l}>\right|^2
\;.\label{u19}
\end{eqnarray}
This is the usual quantum mechanical expression 
which violates Bell's inequality and whose correctness is experimentally
proven. Thus our theory gives the correct expression for the correlation.
Nevertheless, if the joint probability 
(\ref{u18}) exists, it leads to 
\begin{eqnarray}
P(M_1,j,M_2,k)\mbox{\hspace{5.5cm}}\nonumber\\
=\sum_l |c_l|^2 |<\xi(P_1,j)|\phi_{P_1,l}>|^2
|<\xi(P_2,k)|\phi_{P_2,l}>|^2\label{u20}
\end{eqnarray}
which satisfies Bell's inequality and contradicts
Eq.(\ref{u19}). 
Let us demonstrate that no such contradiction appears.
Evidently, the joint probability 
$P(P_1,l_1,P_2,l_2,(0);M_1,j,M_2,k,(t))$ can be physically meaningful 
only if one can compare the initial internal states of $P_1$ and $P_2$
with the final internal states of $M_1$ and $M_2$ by suitable
nondisturbing measurements. 
If we try to compare the initial internal 
states of $P_1$ and of $P_2$ with the final
internal states of $M_1$ and $M_2$, 
the first difficulty appears because we want to compare states
given at different times. Nevertheless, as the initial internal state
of $P_i$ is uniquely related to the final internal state of the
system $P_i+M_i$, the joint probability $P(P_1,l_1,P_2,l_2,(0);M_1,j,M_2,k,(t))$ 
(if exists) coincides with $P(P_1+M_1,l_1,P_2+M_2,l_2,M_1,j,M_2,k)$,
where all the occuring states are given after the measurements.
As the systems $P_1+M_1,\;P_2+M_2,\;M_1,\;M_2$ are not disjointed, 
our {\bf Postulates} do not provide us with an expression for
the joint probability we are seeking for. 
 If we check $|\psi_{M_1}>$ and $|\psi_{M_2}>$ by
 suitable nondisturbing measurements, we destroy
 $|\psi_{P_1+M_1}>$ and $|\psi_{P_2+M_2}>$, inhibiting any comparison.
  On the other hand, if we check
 first $|\psi_{P_1+M_1}>$ and $|\psi_{P_2+M_2}>$, then 
$P(M_1,j,M_2,k)$ changes.
In fact, after suitable measurements 
performed on
$P_i+M_i$ (which do not change the internal states of $P_i+M_i$)
 by further measuring devices $\tilde M_i$ we get for the
internal state  of the whole system
 \begin{eqnarray}
\sum_l c_l\left( \sum_j <\xi(P_1,j)|\phi_{P_1,l}>|\xi(P_1,j)>|m^{(1)}_j>\right)
\nonumber\\
\times\left( \sum_k <\xi(P_2,k)|\phi_{P_2,l}>|\xi(P_2,k)>|m^{(2)}_k>
\right)\nonumber\\
\times|\tilde m^{(1)}_l>|\tilde m^{(2)}_l>\;.\mbox{\hspace{4.3cm}}\label{u21}
\end{eqnarray} 
As the systems $M_1,\;M_2,\;\tilde M_1,\;\tilde M_2$ are disjointed,
we may apply {\bf Postulate C} for $n=4$ and we indeed get for 
$P(\tilde M_1,l_1,\tilde M_2,l_2,M_1,j,M_2,k)$ the expression
(\ref{u18}). Do we get then a contradiction with Eq.(\ref{u19})?
No, because applying {\bf Postulate C} directly for $n=2$, we get in this case
Eq.(\ref{u20}) instead of Eq.(\ref{u19}). Thus we see that the
extra measurements have changed the correlations and our theory 
gives account of this effect consistently.

Summing up, we have seen that the initial internal state
of $P_1$ ($P_2$) determines the first (second) measurement
process, therefore, these states 'carry' the initial correlations
and 'transfer' them to the measuring devices. 
As the measurement processes do not influence each other, 
the observed correlations may stem only from the 'common past'
of the particles.
On the other hand, any attempt to
compare the initial internal states of $P_1$ and $P_2$ with
the results of both measurements changes the correlations,
thus a joint probability for the simultaneous existence of these states
cannot be defined. This means that the reason for the
 violation of Bell's inequality is that the usual derivations
 always assume that the states (or 'stable properties')
 which carry the initial correlations can be freely compared with the results
 of the measurements. This comparability is usually 
 thought to be a consequence of realism.
 According to the present theory, the above assumption
 goes beyond the requirements of realism and proves to be wrong,
 because each of the states $|\psi_{P_1+M_1}>$, 
$|\psi_{P_2+M_2}>$, $|\psi_{M_1}>$ and $|\psi_{M_2}>$ exists individually,
but they cannot be compared without changing the correlations.
\vskip0.5cm
\centerline{\bf Acknowledgements}
\vskip0.5cm
The author is indebted to A.Bringer, G.Eilenberger, M.Eisele, 
R.Graham, G.Gy\"orgyi, F.Haake, H.Lustfeld, P.Rosenqvist,
 P.Sz\'epfalusy, Z.Kaufmann, and G.Vattay for useful discussions 
 and remarks,
 to P.Sz\'epfalusy also for his continued interest in 
 the work and for encouragement.
 The author wants to thank for the hospitality of the {\em Institut f\"ur 
 Festk\"orperphysik, Forschungszentrum J\"ulich GmbH} where a 
 substantial 
 part of the work has been done. 
 This work has been partially supported by the Hungarian Academy of 
 Sciences
 under Grant Nos. OTKA T 017493, OTKA F 17166 and OTKA F 19266.


\begin{thebibliography}{99}
  \bibitem{von Neumann} J.von Neumann, {\em Mathematische Grundlagen 
 der
Quantenmechanik} (Springer, Berlin, 1932).  
 \bibitem{hid} J.S.Bell, Physics {\bf 1} (1964) 195, reprinted
in {\em Proc. of Int. School of Physics ``Enrico Fermi'', Course 49
} ed.: B.d'Espagnat, (Academic Press, New York, 1974).
   \bibitem{exp} A.Aspect, P.Grangier, and G.Roger,  Phys.Rev.Lett. 
 {\bf 47}  (1982) 91, A.Aspect, J.Dalibard, and G.Roger, 
  Phys.Rev.Lett. {\bf 49}  (1982) 1804. 
 \bibitem{Esp} B. d'Espagnat, Sci.Am. {\bf 241} (1979) 128., J.Stat.Phys. {\bf 56} (1989) 747.
 \bibitem{Landau} L.D.Landau and E.M.Lifshitz {\em Quantum mechanics
\hfill\break
[Course on theoretical physics, Vol.3]} (Pergamon Press, New York,
1965) pp.21-24.
\end{thebibliography}
\end{document}